\documentclass[aps,prd,showpacs,nofootinbib,preprintnumbers,onecolumn]{revtex4}

\usepackage{bm}
\usepackage{latexsym}
\usepackage{dcolumn}
\usepackage{amsfonts,amssymb,amsmath}
\usepackage{graphicx,epsfig}
\usepackage{psfrag}
\usepackage{multirow}
  
\newcommand{\beq}{\begin{equation}} 
\newcommand{\eeq}{\end{equation}} 
\newcommand{\bea}{\begin{eqnarray}} 
\newcommand{\eea}{\end{eqnarray}} 
\def\benu{\begin{enumerate}}
\def\eenu{\end{enumerate}}
\def\nn{\nonumber}
\def\nn{\nonumber}

\def\l{\left}
\def\r{\right}
\def\f{\frac}
\def\tx{\tilde x}
\def\dt{\Delta t}
\def\dx{\Delta {\bf x}}
\def\kp{k_{_{\rm P}}}
\def\bkp{{\bar k}_{_{\rm P}}}

\begin{document} 
\title{Modified dispersion relations and the response of\\ 
the rotating Unruh-DeWitt detector}
\author{Sashideep Gutti\footnote{E-mail:~sashideep@hri.res.in}, 
Shailesh Kulkarni\footnote{E-mail:~skulkarni@hri.res.in} and 
L.~Sriramkumar\footnote{E-mail:~sriram@hri.res.in}}
\affiliation{Harish-Chandra Research Institute, Chhatnag Road, Jhunsi, 
Allahabad~211~019, India.}
\date{\today}
\begin{abstract}
We study the response of a rotating monopole detector that is coupled 
to a massless scalar field which is described by a non-linear dispersion 
relation in flat spacetime.
Since it does not seem to be possible to evaluate the response of the 
rotating detector analytically, we resort to numerical computations.
Interestingly, unlike the case of the uniformly accelerated detector
that has been considered recently, we find that defining the transition 
probability rate of the rotating detector poses no difficulties.
Further, we show that the response of the rotating detector can be 
computed {\it exactly}\/ (albeit, numerically) even when it is coupled
to a field that is governed by a non-linear dispersion relation. 
We also discuss the response of the rotating detector in the presence 
of a cylindrical boundary on which the scalar field is constrained to 
vanish.
While super-luminal dispersion relations hardly affect the standard 
results, we find that sub-luminal dispersion relations can lead to 
relatively large modifications.
\end{abstract}
\maketitle


\section{Planck scale modifications, the Unruh effect, and a variant}

Over the last decade or two, there has been intermittent efforts 
towards understanding the possible Planck scale corrections to a 
variety of non-perturbative, quantum field theoretic effects in 
flat as well as curved spacetimes.
While the trans-Planckian nature of the modes that are encountered 
(when the initial conditions are imposed on a quantum field) in the 
context of black hole evaporation and inflationary cosmology have 
cornered most of the attention (see, for instance, Refs.~\cite{tpp-bhe} 
and~\cite{tpp-ic}), a few other problems have been investigated as 
well (see, for example, Refs.~\cite{srini-1998,jacobson-2001,other-e}).
Needless to say, in the absence of a workable theory of quantum gravity, 
it seems imperative to extend such phenomenological analyses to as large 
a set of physical situations as possible (in this context, see 
Ref.~\cite{pqg}, and references therein).

The Unruh effect---viz. the thermal nature of the Minkowski vacuum 
when viewed by an observer in motion along a uniformly accelerated 
trajectory (see any of the following texts~\cite{texts}, or the recent 
review~\cite{crispino-2008})---has certain similarities with Hawking 
radiation from black holes.
Because of this reason, the Unruh effect and its variants provide 
another interesting arena to investigate the quantum gravitational 
effects~\cite{srini-1998,agullo-2008,rinaldi-2009,campo-2010}.

However, due to the lack of a viable theory of quantum gravity, to study
the Planck scale effects, one is forced to consider phenomenological
models constructed by hand.
These models often attempt to capture one or more features expected of 
the actual effective theory obtained by integrating out the gravitational
degrees of freedom. 
An approach that has been extensively considered both in the context of 
black holes and inflationary cosmology are the models that are based on
modified dispersion relations. 
These models effectively introduce a fundamental scale into the theory
by breaking local Lorentz 
invariance~\cite{jacobson-2001,lvm-reviews,raetzel-2010}.
Though it is true that there does not seem to exist any experimental or
observational reason to believe that Lorentz invariance could be violated 
at high energies, theoretically, the hallmark of these models is that they 
often allow quantum field theories to be constructed and calculations to 
be carried out in a consistent and systematic fashion.

In this work, we shall adopt the approach due to the modified dispersion
relations to analyze the Planck scale corrections to the response of the
so-called Unruh-DeWitt detector~\cite{unruh-1976,dewitt-1979}, when it is 
set in motion on a rotating trajectory in flat spacetime~\cite{letaw-1981a,letaw-1981b,paddy-1982,bell-1983-1987,sriram-2002,korsbakken-2004}.
In fact, the case of the uniformly accelerated detector has been studied
recently~\cite{rinaldi-2009}.
The analysis indicates two possible difficulties.
Firstly, evaluating the response of the accelerated detector requires 
resorting to certain approximations to compute the Wightman function
associated with the massless scalar field that is governed by a 
non-linear dispersion relation.
However, the validity of these approximations under which the Wightman 
function can be evaluated in a closed form does not seem to be completely
clear.
Secondly, defining the corresponding transition probability rate of the
accelerating detector poses peculiar problems (for a detailed discussion
on these two points, see Ref.~\cite{campo-2010}).
As we shall see, the rotating trajectory turns out to be a special case 
wherein the second difficulty does not arise; the transition probability 
rate of the rotating detector can be defined in precisely the same 
fashion as in the standard case of the linear dispersion relation.
Further, we find that the first of the two difficulties mentioned above 
can be easily overcome in the case of the rotating trajectory by adopting 
a slightly different method (which avoids having to initially evaluate the 
Wightman function) to calculate the response of the detector.
But, it is known that, even in the conventional case of the linear 
dispersion relation, it is not possible to calculate the response of the 
rotating detector analytically, and one has to resort numerical computations 
to arrive at the transition probability rate~\cite{letaw-1981a}.
We shall illustrate that, the response of the rotating detector can be 
computed {\it exactly},\/ although, numerically, even when the field it 
is coupled to is described by a non-linear dispersion relation. 
In addition, we shall also discuss the response of the rotating detector 
in the presence of a cylindrical boundary on which the scalar field is 
constrained to vanish.
We find that, whereas super-luminal dispersion relations scarcely 
affect the standard results, sub-luminal dispersion relations can 
lead to large modifications.

The plan of the paper is as follows. 
In the following section, after briefly sketching the concept of a
detector, we shall outline as to how the response of the Unruh-DeWitt 
detector can be expressed in terms of the Fourier transform of the 
Wightman function with respect to the differential proper time in the 
frame of the detector.  
We shall then illustrate the transition probability rate of the rotating 
detector in the Minkowski vacuum by numerically computing the integral 
involved. 
In Sec.~\ref{sec:rd-wmdr}, we shall consider the response of the rotating
detector when it is coupled to a massless scalar field that is governed 
by a modified dispersion relation.
However, since it seems difficult to express the Wightman function
corresponding to a scalar field that is described by a non-linear 
dispersion relation in a closed form, we shall adopt another method
to evaluate the response  of the detector.
We shall first re-derive the standard result for the rotating detector 
using the method, and then consider the case wherein the detector is 
coupled to a field that is described by a non-linear dispersion relation.
Importantly, we find that, though we have to resort to a numerical
computation of a particular sum, the response of the detector can be 
evaluated exactly even when it is coupled to a scalar field governed 
by a modified dispersion relation. 
In Sec.~\ref{sec:rd-wb}, we shall discuss the response of a rotating 
detector when the scalar field is assumed to vanish on a cylindrical 
boundary that is located at a radius within the static limit in the 
rotating frame. 
Finally, in Sec.~\ref{sec:discussion}, we shall conclude with a few 
comments on the results of our analysis.

In what follows, we shall set $\hbar=c=1$, and shall work in the
$(3+1)$-dimensional Minkowski spacetime with the metric signature 
of $(+,-,-,-)$.
For convenience, we shall denote the set of four spacetime 
coordinates~$x^{\mu}\equiv(t,{\bf x})$ as~$\tx$.
Also, an overbar shall refer to a suitable dimensionless quantity.
 

\section{The standard response of the rotating Unruh-DeWitt detector}
 
In this section, we shall rapidly summarize the essential aspects of 
the Unruh-DeWitt detector~\cite{unruh-1976,dewitt-1979}, and numerically
compute the standard response of the rotating detector, i.e. when it is 
coupled to a massless scalar field governed by the conventional, linear 
dispersion relation~\cite{letaw-1981a}.

By a detector, one has in mind a point-like object that can be described
by a classical worldline, but which nevertheless possesses internal energy 
levels.
The detectors are essentially described by the interaction Lagrangian for 
the coupling between the internal degrees of freedom of the detector and 
the quantum field.
In our discussion below, we shall consider the quantum field to be a 
massless scalar field, say,~$\phi$.
The Unruh-DeWitt detector is coupled to the scalar field by a monopole 
interaction of the form:~$\l({\cal C}\, \mu(\tau)\, \phi[\tx(\tau)]\r)$, 
where ${\cal C}$ is the coupling constant, $\mu(\tau)$ is the quantity 
that describes the monopole moment of the detector and $\tx(\tau)$ is the 
trajectory of the detector, with $\tau$ being the proper time in the 
detector's frame. 
Upto the first order in the perturbation theory, the amplitude of transition
of the detector from its ground state $\vert E_0\rangle$ (with energy 
$E_{0}$) to an excited state $\vert E\rangle$ (with energy $E$) can easily 
be shown to be~\cite{texts}
\beq
{\cal A}({\cal E})
={\cal M}\; \int\limits_{-\infty}^{\infty}\; d\tau\; 
e^{i\,{\cal E}\, \tau}\;
\langle \Psi\vert\; {\hat \phi}[\tx(\tau)]\;\vert 0\rangle,
\label{eq:ta}
\eeq
where $\mathcal{M}=\l[i\, {\cal C}\, \langle E\vert {\hat \mu}(0)\vert E_{0}
\rangle\r]$, ${\cal E}=\l(E-E_0\r)$, $\vert\Psi\rangle$ is the final 
state of the quantum field, and we have assumed that the field was 
initially in the vacuum state $\vert 0\rangle$. 
Note that the quantity ${\cal M}$ depends only on the internal structure 
of the detector, and not on its motion. 
Therefore, as is the usual practice, we shall drop the quantity hereafter.
The transition probability to all possible final states $\vert \Psi\rangle$ 
of the quantum field is then given by~\cite{texts}
\beq
{\cal P}({\cal E}) =
\sum_{\vert\Psi\rangle}{\vert {\cal A}({\cal E})\vert}^2
= \int\limits_{-\infty}^{\infty}\; d\tau\;
\int\limits_{-\infty}^{\infty}\; d\tau'\; 
e^{-i\,{\cal E}\, (\tau-\tau')}\;
G^{+}\l[\tx(\tau), \tx(\tau')\right],\label{eq:tp} 
\eeq
where $G^{+}\l[\tx(\tau), \tx(\tau')\right]$ denotes the Wightman 
function that is defined as
\beq
G^{+}\left[\tx(\tau), \tx(\tau')\right] 
= \langle 0\vert\, {\hat \phi}[\tx(\tau)]\, 
{\hat \phi}[\tx(\tau')]\,\vert 0 \rangle.
\eeq

When the Wightman function is invariant under time translations in the 
frame of the detector---as it occurs, for instance, in the cases wherein 
the scalar field is described by the conventional dispersion relation 
and the detector is moving along the integral curves of time-like Killing 
vector fields~\cite{letaw-1981a,letaw-1981b,paddy-1982,sriram-2002}---we 
have
\beq
G^{+}\l[\tx(\tau), \tx(\tau')\r]= G^{+}(\tau-\tau').
\eeq
In such situations, the transition probability of the detector 
simplifies to
\beq
{\cal P}({\cal E})
= \lim_{T\to \infty}\;\int\limits_{-T}^{T}\; \f{dv}{2}\; 
\int\limits_{-\infty}^{\infty}\; du\; e^{-i\,{\cal E}\, u}\; G^{+}(u), 
\eeq
where 
\beq
u=(\tau-\tau')\quad{\rm and}\quad v=(\tau+\tau').\label{eq:uv}
\eeq
The above expression then allows us to define the transition probability 
{\it rate}\/ of the detector to be~\cite{texts}
\bea
{\cal R}({\cal E}) 
= \lim_{T\to \infty}\; \l[{\cal P}({\cal E})/T\r]
= \int\limits_{-\infty}^\infty\; du\; e^{-i\,{\cal E}\, u}\; 
G^{+}(u).\label{eq:tpr}
\eea
When the massless scalar field is described by the standard, linear 
dispersion relation, the Wightman function $G^{+}\l(\tx, \tx'\r)$ in
the Minkowski vacuum is given by~\cite{texts}
\beq
G^+\l(\tx,\tx'\r)
=-\l(\frac{1}{4\, \pi^2}\r)\,
\l(\frac{1}{(\dt-i\,\epsilon)^2
-\dx^2}\r),\label{eq:mgfn-sc}
\eeq
where $\dt = \l(t - t'\r)$, $\dx=\l({\bf x}-{\bf x}'\r)$, 
and $\epsilon\to 0^{+}$.
Given a trajectory $\tx(\tau)$, the response of the detector is obtained 
by substituting the trajectory in this Wightman function and evaluating 
the transition probability rate~(\ref{eq:tpr}).

\begin{figure}[!htb]
\vskip 15pt
\begin{center}
\resizebox{330pt}{220pt}{\includegraphics{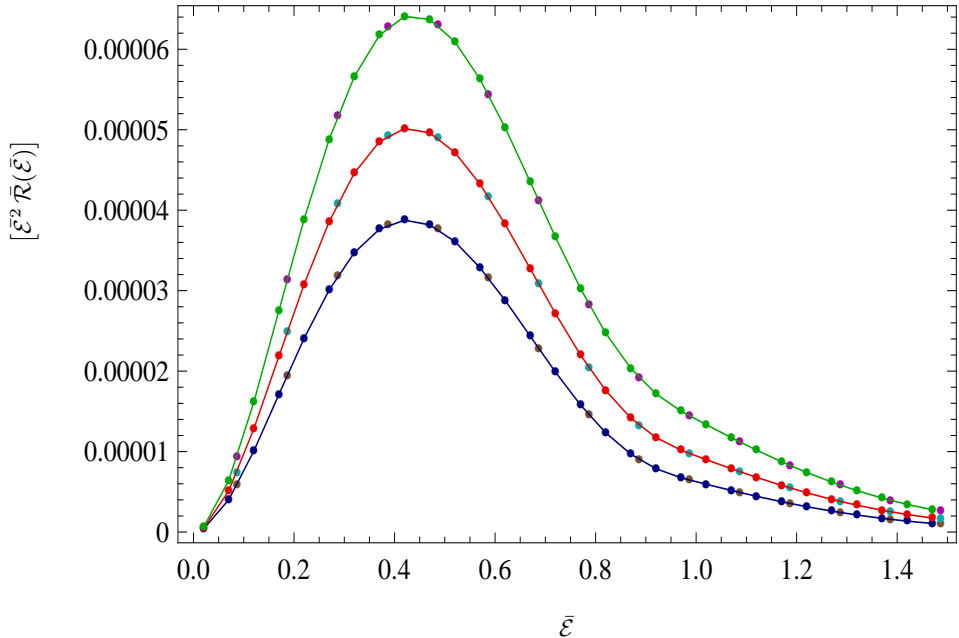}}
\vskip -170 pt \hskip -375 pt 
\rotatebox{90}{$\l[{\bar {\cal E}}^2\, 
{\bar {\cal R}}({\bar {\cal E}})\r]$}
\vskip 138 pt
\hskip 35 pt ${\bar {\cal E}}$ 
\vskip 5 pt
\caption{The transition probability rate of the rotating Unruh-DeWitt 
detector that is coupled to a massless scalar field governed by the 
standard dispersion relation.
The blue, the red and the green dots denote the numerical results
obtained through the computation of the integral~(\ref{eq:tpr}) along 
the rotating trajectory, and they correspond to the following three
choices of the quantity $(\sigma\, \Omega)=0.325$, $0.350$ and $0.375$, 
respectively. 
The solid curves simply join the dots to guide the eye.
The intervening dots of an alternate color that appear on these curves 
denote the corresponding numerical results that have been arrived at 
by another method as described in the following section [they actually 
correspond to the sum~(\ref{eq:tpr-sc-rt-smfv})].
It is evident that the results from the two methods match very well.}
\label{fig:rd-sc}
\end{center}
\end{figure}

The time-like Killing vector field $\xi^{\mu}$ that generates the rotational 
motion is given by~\cite{letaw-1981a,letaw-1981b,paddy-1982,sriram-2002}
\beq
\xi^{\mu}=(\gamma,-\lambda\, y,\, \lambda\,x,\,0). 
\eeq
The integral curve corresponding to this Killing vector can be expressed 
in terms of the proper time, say, $\tau$, as follows:
\beq
\tx(\tau)=\l[(\gamma\,\tau),\,
\sigma\, \cos\,(\gamma\,\Omega\,\tau),\,
\sigma\, \sin\, (\gamma\,\Omega\,\tau),\,0\r],\label{eq:rt}
\eeq
where we have set $\lambda=\l(\gamma\, \Omega\r)$. 
The constants $\sigma$ and $\Omega$ denote the radius of the circular path 
along which the detector is moving and the angular velocity of the detector, 
respectively.
The quantity $\gamma=\l[1-(\sigma\, \Omega)^2\r]^{-1/2}$ is the Lorentz 
factor that relates the Minkowski time to the proper time in the frame 
of the detector.
For such a rotational motion, we find that
\bea
\dt^{2} &=& \gamma^2\, \l(\tau - \tau'\r)^2 
=\gamma^2\, u^2,\label{eq:rtdtsq}\\
\dx^{2} &=& 4\,\sigma^{2}\, \sin^{2}\,[\gamma\,\Omega\, (\tau-\tau')/2]
=4\,\sigma^{2}\, \sin^{2}\,(\gamma\,\Omega\, u/2).
\label{eq:rtdxsq}
\eea
Upon substituting these quantities in the expression~(\ref{eq:mgfn-sc}),
the Wightman function along the rotating trajectory can be obtained to be
\beq
G^+\l[\tx(\tau),\tx(\tau')\r]
=G^+(u)
=-\l(\frac{1}{4\,\pi^2\, \sigma^{2}}\r)\,
\l(\frac{1}{(\gamma/\sigma)^2\; (u-i\,\epsilon)^2
-4\,\sin^2\,(\gamma\, \Omega\, u/2)}\r).
\label{eq:mgfn-sc-rt}
\eeq
However, unfortunately, it does not seem to be possible to evaluate the 
corresponding transition probability rate ${\cal R}({\cal E})$ analytically. 
We have arrived at the response of the rotating detector by substituting the 
Wightman function~(\ref{eq:mgfn-sc-rt}) in the expression~(\ref{eq:tpr}), 
and numerically computing the integral involved.
If we define the dimensionless energy to be ${\bar {\cal E}}=({\cal E}/
\gamma\, \Omega)$, we find that the dimensionless transition probability 
rate ${\bar {\cal R}}({\bar {\cal E}})\equiv\l[\sigma\, {\cal R}({\bar
{\cal E}})\r]$ of the detector depends only on the dimensionless quantity 
$(\sigma\, \Omega)$ that describes the linear velocity of the detector.
In Fig.~\ref{fig:rd-sc}, we have plotted the transition probability rate 
of the detector for three different values of the quantity $(\sigma\, 
\Omega)$~\cite{letaw-1981a}.
We should mention here that, in order to check the accuracy of the 
numerical procedure that we have used to evaluate the 
integral~(\ref{eq:tpr}) for the rotating trajectory, we have compared 
the results from the numerical code with the analytical ones that are 
available for the case of the uniformly accelerated 
detector~\cite{unruh-1976,dewitt-1979,letaw-1981a,paddy-1982,sriram-2002}. 
This comparison clearly indicates that the numerical procedure we have
adopted to evaluate the integral~(\ref{eq:tpr}) is quite accurate (for 
details, see App.~A).


\section{Response of the rotating detector coupled to a scalar field
governed by a modified dispersion relation}\label{sec:rd-wmdr}

In this section, we shall calculate the response of the rotating 
detector when it is coupled to a massless, scalar field that is 
governed by a modified dispersion relation of the following form:
\beq
\omega=k\, \l[1+\alpha\, \l(k/\kp\r)^{2}
+\beta\, \l(k/\kp\r)^{4}\r]^{1/2},\label{eq:mdr}
\eeq
where $\omega$ is the frequency corresponding to the mode ${\bf k}$, 
$k=\vert {\bf k}\vert$, $\kp$ denotes the fundamental scale beyond 
which the modifications to the linear dispersion relation become 
important, while $\alpha$ and $\beta$ are dimensionless constants whose 
magnitudes are of order unity. 
Note that the above dispersion relation is super-luminal or sub-luminal 
depending upon whether the constants $\alpha$ and $\beta$ are positive 
or negative.
Evidently, if we can evaluate Wightman function associated with the
scalar field described by the non-linear dispersion relation~(\ref{eq:mdr}),
we may then be able to evaluate the corresponding transition probability
rate of the detector as we did the previous section. 
However, unlike the standard case, it turns out to be difficult to 
compute the Wightman function of such a scalar field exactly.
As a result, one has to either take recourse to some approximations to 
arrive at an analytic expression for the Wightman function or adopt
another method to evaluate the response of the detector.
Before going on to calculate the response of the rotating detector 
through another method, we shall briefly discuss the approximations 
that are typically made in the evaluation of the Wightman function 
for the quadratic dispersion relation (i.e. the case wherein 
$\beta=0$)~\cite{rinaldi-2009}, and the difficulties that can be 
encountered in defining the corresponding transition probability 
rate~\cite{campo-2010}.

The equation of motion of the scalar field $\phi$ that is described by 
the dispersion relation~(\ref{eq:mdr}) is given by
\beq
\Box\,\phi
+\l(\frac{\alpha}{\kp^2}\r)\, \nabla^2\,\l(\nabla^2\,\phi\r)
-\l(\frac{\beta}{\kp^4}\r)\, 
\nabla^2\,\l[\nabla^2\,\l(\nabla^2\,\phi\r)\r]=0,
\eeq
where $\Box$ is the D'Alembertian corresponding to the four dimensional
Minkowski spacetime, while $\nabla^2$ is the three dimensional, spatial 
Laplacian.
Whereas the first term in the above equation is the standard one, the 
remaining terms arise due to the non-linearity in the dispersion relation.
Such terms can be generated by adding suitable terms to the original action 
describing the scalar field~\cite{jacobson-2001,lvm-reviews}.
These additional terms preserve rotational invariance, but break Lorentz
invariance and, in fact, this property is common to all the theories that
are described by a non-linear dispersion relation.
The normal modes of the scalar field in flat spacetime remain plane waves 
as in the standard case (with the frequency and the wavenumber related by 
the modified dispersion relation), and the quantization of the field can 
be carried out in the same fashion.
In the Minkowski vacuum, the Wightman function for any such field can 
be expressed as (see, for example, Refs.~\cite{jacobson-2001})
\beq
G^{+}_{_{\rm M}}(\tx,\tx')
=\int \frac{d^{3}{\bf k}}{(2\,\pi)^3\, (2\, \omega)}\, 
e^{-i\,\omega\, (t-t')}\, e^{i\,{\bf k}\cdot\, ({\bf x}-{\bf x}')}
\label{eq:mgfn-mdr}
\eeq
with $\omega$ being related to $k=\vert {\bf k}\vert$ by the given 
non-linear dispersion relation.

As we have said before, it does not seem to be possible to calculate 
the above integral exactly for a modified dispersion relation of the
form~(\ref{eq:mdr}). 
(In fact, we are not aware of any non-linear dispersion relation for 
which the Wightman function can be evaluated exactly in a closed form.)
To arrive at an analytic expression for the Wightman function, two
approximations are resorted to. 
One first expands the quantity $\omega$ in the denominator and in the
exponential, say, upto the first order in the inverse power of $\kp^2$. 
Then, the exponential term containing the correction is expanded upto to 
the same order in $\kp^{-2}$ as well.
Evidently, such an expansion will be valid only for $k<\kp$.
So, a cut-off at large momentum (at $k \simeq \kp$) is further assumed in 
order to carry out the integral over $k$. 
Under these conditions, for the case wherein $\beta=0$, it can be 
shown that the resulting Wightman function consists of two terms, 
with the leading term being the standard, Lorentz invariant, Wightman
function~(\ref{eq:mgfn-sc}), whereas the correction is given 
by~\cite{rinaldi-2009,campo-2010}
\beq
G^{+}_{_{\rm C}}(\tx,\tx')
=-\l(\frac{\alpha}{4\,\pi^2\, \kp^2}\r)\;
\l(\frac{15\, \dt^4
+10\, \dt^2\, \dx^2-\dx^4}{[(\dt - i\,\epsilon)^2-\dx^2]^4}\r).
\label{eq:mgfn-ct}
\eeq

It is important to note that, as opposed to the leading term, the above 
correction $G^+_{_{\rm C}}(\tx, \tx')$ is {\it not}\/ Lorentz invariant.
Due to the lack of Lorentz invariance, unlike the original Wightman 
function, the quantity $G^+_{_{\rm C}}[\tx(\tau), \tx(\tau')]$ may 
not be invariant under time translations in the frame of the detector, 
even if it is moving along the integral curves of time-like Killing 
vector fields, such as, for instance, the popular, uniformly accelerated
trajectory~\cite{rinaldi-2009,campo-2010}. 
In other words, generically, $G^+_{_{\rm C}}[\tx(\tau),\tx(\tau')]$ 
will be a function of $u$ {\it as well as}\/ $v$ along the non-inertial
trajectory under consideration.
Since the transition probability of the detector is linear in the 
Wightman function, clearly, the response of the detector will be 
the sum of the standard response, and a term that is proportional 
to $\kp^{-2}$, given by
\beq
{\cal P}_{_{\rm C}}({\cal E})
=\int\limits_{-\infty}^{\infty}\; d\tau
\int\limits_{-\infty}^{\infty}\; d\tau'\;
e^{-i\, {\cal E}\, \l(\tau-\tau'\r)}\;
G^{+}_{_{\rm C}}[\tx(\tau),\tx(\tau')].\label{eq:tp-c}
\eeq 
However, because of the dependence of the term 
$G^+_{_{\rm C}}[\tx(\tau),\tx(\tau')]$ on $v$ as well, in general, it 
turns out to be more involved to define the corresponding transition 
probability rate (for a detailed discussion in this context, see 
Ref.~\cite{campo-2010}).

Recall that, along the rotating trajectory, the quantities $\dt$ and $\dx$
depend only on $u$ [cf.~Eqs.~(\ref{eq:rtdtsq}) and (\ref{eq:rtdxsq})].
Therefore, the rotating trajectory turns out to be a special case where, as 
in the original Wightman function, the correction term 
$G^+_{_{\rm C}}[\tx(\tau),\tx(\tau')]$ proves to be a function only of $u$, 
and not of $v$.
This feature essentially arises due to the fact that the term 
$G^+_{_{\rm C}}(\tx,\tx')$, though it is not Lorentz invariant, 
preserves rotational invariance.
Hence, we can define the corresponding transition probability rate 
just as in the standard case.
However, some concerns have been raised about the validity 
of the approximations that have been made to arrive at the above 
form for $G^+_{_{\rm C}}(\tx,\tx')$~\cite{campo-2010}.
Therefore, we shall not use the term~(\ref{eq:mgfn-ct}) to evaluate the
corrections to the response of the rotating detector, as has been done 
recently in the uniformly accelerated case~\cite{rinaldi-2009}.
Instead, in what follows, we shall compute the Planck scale modifications 
to the response of the rotating detector by adopting a slightly different 
method to compute the transition probability rate.
As we shall illustrate, the method allows us to evaluate the response of
the detector {\it exactly},\/ even for the case of a scalar field that is
described by a modified dispersion relation.
 

\subsection{Another method to evaluate the standard response of the 
rotating detector}

In the previous section, we had obtained the standard response of the 
rotating detector by evaluating the Fourier transform of the Wightman 
function with respect to the differential proper time in the frame of 
the detector.
Here, we shall firstly re-derive the result in a slightly different 
fashion, a method which, in retrospect, would seem evident in the 
rotating case.
It essentially involves expressing the Wightman function as a sum over
the normal modes, and first evaluating the integral over the differential
proper time $u$ before computing the sum.
As we shall see, the method can be extended in a straightforward manner
to the case wherein the scalar field is described by a modified dispersion
relation.

To start with, we shall work in the cylindrical polar coordinates, say, 
$(t,\rho,\theta,z)$, instead of the cartesian coordinates, as they turn 
out to be more convenient.
In terms of the cylindrical coordinates, the trajectory~(\ref{eq:rt}) of 
the rotating detector can be written in terms of the proper time $\tau$
as follows:
\beq
\tx(\tau)=\l[(\gamma\, \tau),\,\sigma,\,(\gamma\, \Omega\, \tau),\,0\r].
\label{eq:rt-cyc}
\eeq
Using certain well-established properties of the Bessel functions, it can 
be easily shown that, along the trajectory of the rotating detector, the 
standard Minkowski Wightman function~(\ref{eq:mgfn-sc}) can be expressed 
as 
\beq
G^{+}[\tx(\tau),\tx(\tau')]=G^{+}(u)
=\sum^{\infty}_{m=-\infty}\,
\int\limits^\infty_0\frac{dq\, q}{\l(2\,\pi\r)^2}\,
\int\limits^\infty_{-\infty}\frac{dk_z}{\l(2\,\omega\r)}\,
J_{m}^{2}(q\,\sigma)\; e^{-i\,\gamma\,\l(\omega-m\,\Omega\r)\, u},
\label{eq:mgfn-rt-ms}
\end{equation} 
where $J_{m}(q\, \sigma)$ denote the Bessel functions of order $m$, with 
$\omega$ given by
\beq
\omega=\l(q^{2}+k_{z}^{2}\r)^{1/2}.
\eeq
The corresponding transition probability rate of the rotating detector
is then given by
\beq
{\cal R}({\cal E})
=\sum^{\infty}_{m=-\infty}\,
\int\limits^\infty_0\frac{dq\, q}{\l(2\,\pi\r)}\;
\int\limits^\infty_{-\infty}\;\frac{dk_z}{\l(2\,\omega\r)}\,
J_{m}^{2}(q\,\sigma)\; 
\delta^{(1)}\l[{\cal E}+\gamma\, \l(\omega-m\,\Omega\r)\r].
\label{eq:tp-cc}
\eeq
Since, ${\cal E}>0$, $\omega\ge 0$ (as is appropriate for positive 
frequency modes), and $\Omega$ too is a positive definite quantity 
by assumption, the delta function in the above expression will be 
non-zero only when $m \ge {\bar {\cal E}}$, where ${\bar {\cal E}}=
\l[{\cal E}/\l(\gamma\, \Omega\r)\r]$ is the
dimensionless energy.
Hence, the response of the detector can be expressed as
\beq
{\cal R}({\bar {\cal E}})
=\sum^{\infty}_{m\ge {\bar {\cal E}}}\,
\int\limits^\infty_0\frac{dq\, q}{\l(2\,\pi\r)}\;
\int\limits^\infty_{-\infty}\frac{dk_z}{\l(2\,\omega\r)}\;
J_{m}^{2}(q\,\sigma)\; 
\l[\f{\delta^{(1)}(k_z-\kappa_{z})}{\gamma\,
\l|\l(d\omega/dk_z\r)\r|_{\kappa_z}}\r],\label{eq:tpr-sc-rt-msev}
\end{equation}
where $\kappa_{z}$ are the two roots of $k_{z}$ from the following 
equation: 
\beq
\omega= \l(m-{\bar {\cal E}}\r)\, \Omega.\label{eq:r-eq}
\eeq
The roots are given by
\beq
\kappa_z=\pm\l(\delta^{2}-q^2\r)^{1/2},\label{eq:r-kzb-sc}
\eeq
where, for convenience, we have set 
\beq
\delta=\l({\bar \delta}\, \Omega\r)
=\l(m-{\bar {\cal E}}\r)\, \Omega.\label{eq:delta}
\eeq
As both the positive and negative roots of $\kappa_{z}$ contribute 
equally, we obtain the dimensionless transition probability rate of 
the rotating detector to be
\beq
{\bar {\cal R}}({\bar {\cal E}})\equiv
\sigma\, {\cal R}({\bar {\cal E}})
=\l(\f{\sigma}{2\,\pi\,\gamma}\r)\
\sum^{\infty}_{m\ge {\bar {\cal E}}}\,
\int\limits_{0}^{\delta}\; dq\; q\;
\l(\f{J_{m}^{2}(q\,\sigma)}{\l(\delta^{2}-q^{2}\r)^{1/2}}\r),
\eeq
where we have set the upper limit on $q$ to be $\delta$ since $\kappa_{z}$ 
is a real quantity [cf.~Eq.~(\ref{eq:r-kzb-sc})].
We find that the integral over~$q$ can be expressed in terms of the 
hypergeometric functions (see, for instance, Ref.~\cite{prudnikov-1986}).
Therefore, the transition probability rate of the rotating detector can
be written as 
\beq
{\bar {\cal R}}({\bar {\cal E}})
=\l(\f{1}{2\,\pi\,\gamma}\r)\
\sum^{\infty}_{m\ge {\bar {\cal E}}}\,
\l(\f{\l(\sigma\,\Omega\, 
{\bar \delta}\r)^{(2\,m+1)}}{\Gamma\l(2\,m+2\r)}\r)\; 
_1F_2\l[[m+(1/2)];[m+(3/2)],(2\,m+1);
-\l(\sigma\,\Omega\, {\bar \delta}\r)^2\r],
\label{eq:tpr-sc-rt-smfv}
\eeq
where $_1F_2\l(a;b,c;x\r)$ denotes the hypergeometric function, while 
$\Gamma(x)$ is the usual Gamma function.
It does not seem to be possible to arrive at a closed form expression for
this sum, but the sum converges very quickly, and hence proves to be easy 
to evaluate numerically.
In Fig.~\ref{fig:rd-sc}, we have plotted the numerical results for the
above sum for the same values of the linear velocity $(\sigma\, \Omega)$ 
that we had plotted the results obtained from Fourier transforming the 
Wightman function~(\ref{eq:mgfn-sc-rt}) along the rotating trajectory.
It is clear from the figure that the results from the two different methods
match each other rather well. 


\subsection{The case of a scalar field described by a quadratic 
dispersion relation}

When the scalar field is governed by a modified dispersion relation, 
using the expression~(\ref{eq:mgfn-mdr}) for the corresponding Wightman 
function, it is straightforward to show that, along the rotating trajectory, 
the function can be expressed exactly as in Eq.~(\ref{eq:mgfn-rt-ms}),
with the frequency $\omega$ being related to the wavenumbers $q$ and 
$k_{z}$ by the non-linear dispersion relation.
Clearly, in such a case, the corresponding transition probability rate of
the detector will again be given by Eq.~(\ref{eq:tpr-sc-rt-msev}) with 
$\omega$ suitably defined.
In fact, we should stress here that the result is actually valid for 
{\it any}\/ non-linear dispersion relation.

Let us now consider the response of the rotating detector for the 
dispersion relation~(\ref{eq:mdr}), but {\it with $\beta$ set to 
zero}.\/
In such a case, $\omega$ is related to the wavenumbers $q$ and $k_{z}$ 
as follows:
\beq
\omega
=\l(q^{2}+k_{z}^{2}\r)^{1/2}\; 
\l[1 + \l(\f{\alpha}{\kp^2}\r)\,\l(q^{2}+k_{z}^{2}\r)\r] ^{1/2}.
\eeq
Also, we find that the roots $\kappa_{z}$ [from Eq.~(\ref{eq:r-eq})] 
are given by
\beq
\kappa_{z}^2= \pm \l(\frac{\kp^2}{2\,\alpha}\r)\,
\l[1+\l(\f{4\,\alpha\,\delta^2}{\kp^2}\r)\r]^{1/2}
-\l(\f{\kp^2}{2\,\alpha}\r)-q^2,
\eeq
with $\delta$ defined as in Eq.~(\ref{eq:delta}).
Note that $\kappa_{z}^{2}$ has to be positive definite, since 
$\kappa_{z}$ is a real quantity. 

Let us first consider the case when $\alpha$ is positive.
When, say, $\alpha=1$, the two roots that contribute to the delta 
function in Eq.~(\ref{eq:tpr-sc-rt-msev}) can be written as
\beq
\kappa_{z}=\pm\l(\delta_{+}^{2}-q^2\r)^{1/2},
\eeq
where $\delta_{+}^{2}$ is given by the expression
\beq
\delta_{+}^{2}=\l(\frac{\kp^2}{2}\r)\, 
\l(\l[1+\l(\f{4\,\delta^2}{\kp^2}\r)\r]^{1/2}-1\r)
=\l(\frac{{\bar \delta}_{+}^{2}}{\sigma^{2}}\r)
=\l(\frac{{\bkp}^2}{2\, \sigma^{2}}\r)\, 
\l(\l[1+\l(\f{4\,(\sigma\,\Omega\, 
{\bar \delta})^2}{\bkp^2}\r)\r]^{1/2}-1\r),
\label{eq:delta-plus}
\eeq
$\bkp=(\sigma\, \kp)$ denotes the dimensionless fundamental 
scale, and the sub-script in $\delta_{+}$ refers to the fact 
that we are considering a super-luminal dispersion relation.
Further, as $\kappa_{z}$ is real, we require that 
$q\le  \delta_{+}$.
As in the standard case, the positive and negative roots of 
$\kappa_{z}$ above contribute equally.
Therefore, the response of the rotating detector is given by
\beq
{\bar {\cal R}}_{_{\rm M}}({\bar {\cal E}})
=\sigma\, {\cal R}_{_{\rm M}}({\bar {\cal E}})
=\l(\f{\sigma}{2\,\pi\,\gamma}\r)\
\sum^{\infty}_{m\ge {\bar {\cal E}}}\, 
\l[1+\l(\f{2\,\delta_{+}^2}{\kp^2}\r)\r]^{-1}\;
\int\limits_{0}^{\delta_{+}}\; dq\; q\;
\l(\f{J_{m}^{2}(q\,\sigma)}{\l(\delta_{+}^{2}-q^{2}\r)^{1/2}}\r),
\eeq
and the integral over $q$ can be carried out as earlier to arrive 
at the result
\bea
{\bar {\cal R}}_{_{\rm M}}({\bar {\cal E}})
&=&\l(\f{1}{2\,\pi\,\gamma}\r)\
\sum^{\infty}_{m\ge {\bar {\cal E}}}\,
\l(\f{{\bar \delta}_{+}^{(2\,m+1)}}{\Gamma\l(2\,m+2\r)}\r)\; 
\l[1+\l(\frac{2\,{{\bar \delta}_{+}}^2}{\bkp^2}\r)\r]^{-1}\,
_1F_2\l[[m+(1/2)];[m+(3/2)],(2\,m+1);
-{\bar \delta}_{+}^2\r].\qquad
\label{eq:tpr-rd-psm-spldr}
\eea
It should be stressed that this expression for the transition probability 
rate is exact.

Let us now turn to understanding the behavior of the above 
transition probability rate for large~$\bkp$.
It is clear that, as $\bkp\to \infty$, ${\bar \delta}_{+}\to
(\sigma\, \Omega\, {\bar \delta})$ and, hence, the transition 
transition probability rate~(\ref{eq:tpr-rd-psm-spldr}) reduces 
to the expression that we had arrived at earlier for the standard
dispersion relation [viz.~Eq.~(\ref{eq:tpr-sc-rt-smfv})], as required.
Let us now expand the transition probability 
rate~(\ref{eq:tpr-rd-psm-spldr}) retaining terms upto 
${\cal O}[\l(\delta/\kp\r)^{2}]$. 
Note that, in such a case, $\delta_{+}$ reduces to
\beq
\delta_{+} 
\simeq
\delta\, \left[1 - \l(\frac{\delta^{2}}{2\,\kp^{2}}\right)\r],
\eeq 
so that we have
\beq
\delta_+^{(2\,m+1)} 
\simeq \delta^{(2\,m+1)}
-(2\,m+1)\, \l(\frac{\delta^{(2\, m+3)}}{2\, \kp^2}\r)
\quad{\rm and}\quad
\l[1+\l(\frac{2\,\delta_{+}^2}{k_{_{\rm P}}^2}\r)\r]^{-1} 
\simeq 
1-\l(\frac{2\,\delta^2}{k_{_{\rm P}}^2}\r).
\eeq
Moreover, in the limit of our interest, the hypergeometric function 
in Eq.~(\ref{eq:tpr-rd-psm-spldr}) can be written as 
\bea
& &\!\!\!\!\!\!\!\!\!\!\!\!\!\!\!\!\!\!\!\!
_1F_2\l[[m+(1/2)];[m+(3/2)],(2\,m+1);
-{\bar \delta}_{+}^2\r]\nn\\
& &\!\!\!\!\!\!\!\!
\simeq\, 
_1F_2\l[[m+(1/2)];[m+(3/2)],(2\,m+1);
-\l(\sigma\,\Omega\, {\bar \delta}\r)^2\r]\nn\\
& &\qquad 
+ \l(\f{(\sigma\, \Omega\,{\bar \delta})^{2}}{\bkp^2}\r)\,
\l(\f{[m+(1/2)]\; (\sigma\, \Omega\,{\bar \delta})^{2}}{[m+(3/2)]\; 
(2\, m+1)}\r)\;
_1F_2\l[[m+(3/2)];[m+(5/2)],(2\,m +2);
-\l(\sigma\,\Omega\, {\bar \delta}\r)^2\r].
\eea
Upon using the above expansions, we obtain the response of the 
detector at ${\cal O}[\l(\delta/\kp\r)^{2}]$ to be
\bea
\!\!\!\!\!\!\!\!\!\!\!\!\!\!\!\!\!\!\!\!\!\!\!\!
{\bar {\cal R}}_{_{\rm M}}({\bar {\cal E}})
&\simeq&\l(\f{1}{2\,\pi\,\gamma}\r)
\sum^{\infty}_{m\ge {\bar {\cal E}}}\,
\l(\f{\l(\sigma\,\Omega\, 
{\bar \delta}\r)^{(2\,m+1)}}{\Gamma\l(2\,m+2\r)}\r)\, 
_1F_2\l[[m+(1/2)];[m+(3/2)],(2\,m+1);
-\l(\sigma\,\Omega\, {\bar \delta}\r)^2\r]\nn\\
& & - \, 
\l(\f{1}{2\,\pi\,\gamma}\r)\,
\l(\f{(\sigma\, \Omega\,{\bar \delta})^{2}}{\bkp^2}\r)\,
\sum^{\infty}_{m\ge {\bar {\cal E}}}\,
\l(\f{[m+(5/2)]\,
\l(\sigma\,\Omega\,{\bar \delta}\r)^{(2\,m+1)}}{\Gamma\l(2\, 
m+2\r)}\r)\nn\\
& &\qquad\qquad\qquad\qquad\qquad\qquad\qquad\qquad\times\,
_1F_2\l[[m+(1/2)];[m+(3/2)],(2\,m+1);
-\l(\sigma\,\Omega\,{\bar \delta}\r)^2\r]\nn\\
& & + \, \l(\f{1}{2\,\pi\,\gamma}\r)\,
\l(\f{(\sigma\, \Omega\,{\bar \delta})^{2}}{\bkp^2}\r)\,
\sum^{\infty}_{m\ge {\bar {\cal E}}}\,
\l(\f{[m + (1/2)]\; \l(\sigma\,\Omega\,
{\bar \delta}\r)^{(2\,m+3)}}{[m+(3/2)]\, (2\,m+1)\;
\Gamma\l(2\,m+2\r)}\r)\nonumber\\
& & \qquad\qquad\qquad\qquad\qquad\qquad\qquad\qquad\times\;
 _1F_2\l[[m+(3/2)];[m+(5/2)],(2\,m+2);
-\l(\sigma\,\Omega\,{\bar \delta}\r)^2\r]. 
\eea 
Evidently, the first term in this expression corresponds to the
conventional transition probability rate 
[cf. Eq.~(\ref{eq:tpr-sc-rt-smfv})], while the other two terms 
represent the leading corrections to the standard result.
  
Let us now turn to considering the sub-luminal dispersion relation. 
When $\alpha$ is negative, say, $\alpha=-1$ (and $\beta$ is again zero), 
the roots $\kappa_{z}$ are given by
\beq
\kappa_{z}=\pm\l(\delta_{-}^{2}-q^2\r)^{1/2}
\eeq
with $\delta_{-}^{2}$ defined as
\beq
(\delta_{-}^{\pm})^{2}
=\l(\frac{\kp^2}{2}\r)\, 
\l(1\pm\l[1-\l(\f{4\,\delta^2}{\kp^2}\r)\r]^{1/2}\r)
=\l(\f{({\bar \delta}_{-}^{\pm})^{2}}{\sigma^{2}}\r)
=\l(\frac{\bkp^2}{2\, \sigma^2}\r)\, 
\l(1\pm\l[1-\l(\f{4\,(\sigma\, \Omega\,
{\bar \delta})^2}{\bkp^2}\r)\r]^{1/2}\r),
\eeq
where the minus sign in the sub-script represents that it corresponds 
to the sub-luminal case (i.e. when $\alpha$ is negative), while the 
super-scripts denote the two different possibilities of $\delta_{-}$.  
As in the super-luminal case (i.e. when $\alpha=1$), we require $q\le 
\delta_{-}^{\pm}$, if $\kappa_{z}$ is to remain real.
Moreover, note that, unlike the super-luminal case, there also arises
an upper limit on the sum over $m$.
We require that $\delta\leq(\kp/2)$, in order to ensure that 
$\delta_{-}^{\pm}$ is real.
This corresponds to $m\le \l[{\bar {\cal E}}
+(\bkp/2\,\sigma\, \Omega)\r]$.
Therefore, for the sub-luminal dispersion relation, we find that we 
can write the response of the rotating detector as follows:
\bea
\!\!\!\!\!\!\!\!\!\!\!\!
{\bar {\cal R}}_{_{\rm M}}({\bar {\cal E}})
&=&\l(\f{1}{2\,\pi\,\gamma}\r)\,
\sum^{{\bar {\cal E}}+(\bkp/2\,\sigma\,\Omega)}_{m\ge {\bar {\cal E}}}\,
\l(\f{\l({\bar \delta}_{-}^{-}\r)^{(2\,m+1)}}{\Gamma(2\,m+2)}\r)\,
\l[\;\l\vert\, 1-\l(\f{2\,({\bar \delta}_{-}^{-})^2}{\bkp^2}\r)
\r\vert\;\r]^{-1}\nn\\ 
& &\qquad\qquad\qquad\qquad\qquad\qquad\qquad\qquad\quad\times\;
_1F_2\l[[m+(1/2)];[m+(3/2)],(2\,m+1);
-\l({\bar \delta}_{-}^{-}\r)^2\r]\nn\\
& &
+\,\l(\f{1}{2\,\pi\,\gamma}\r)\
\sum^{{\bar {\cal E}}
+(\kp/2\,\sigma\,\Omega)}_{m\ge {\bar {\cal E}}}\,
\l(\f{\l({\bar \delta}_{-}^{+}\r)^{(2\,m+1)}}{\Gamma\l(2\,m+2\r)}\r)\,
\l[\;\l\vert\, 1-\l(\frac{2\,
({\bar \delta}_{-}^{+})^2}{\bkp^2}\r)\r\vert\;\r]^{-1}\nn\\ 
& &\qquad\qquad\qquad\qquad\qquad\qquad\qquad\qquad\quad\times\;
_1F_2\l[[m+(1/2)];[m+(3/2)],(2\,m+1);
-\l({\bar \delta}_{-}^{+}\r)^2\r].
\label{eq:tpr-rd-psm-sbldr}
\eea

The origin of the upper limit on $m$ as well as the second term in 
the above expression for the response of the rotating detector can 
be easily understood. 
In the case of the super-luminal dispersion relation, $\omega$ is 
a monotonically increasing function of $q$ and $k_{z}$. 
So, there exist only two real roots of $k_{z}$ corresponding to a 
given $\omega$. 
Also, $\omega^{2}$ remains positive definite for all the modes.
But, in the sub-luminal case, after a rise, $\omega$ begins to decrease 
for sufficiently large values of $q$ and $k_{z}$.
In fact, $\omega^{2}$ even turns negative at a suitably large 
value~\cite{jacobson-2001}. 
It is this feature of the sub-luminal dispersion relation which leads 
to the upper limit on $m$.
(The upper limit ensures that we avoid complex frequencies. 
Such a cut-off can be achieved if we assume that, say, the detector
is not coupled to modes with $m$ beyond a certain value, when the 
frequency turns complex.)
The additional two roots of $k_{z}$ that contribute to the detector 
response in the sub-luminal case arise as a result of the decreasing 
$\omega$ at large $q$ and $k_{z}$.
The second term in the above transition probability rate of the rotating
detector corresponds to the contributions from these two extra roots.

\begin{figure}[!htb]
\vskip 15pt
\begin{center}
\resizebox{330pt}{220pt}{\includegraphics{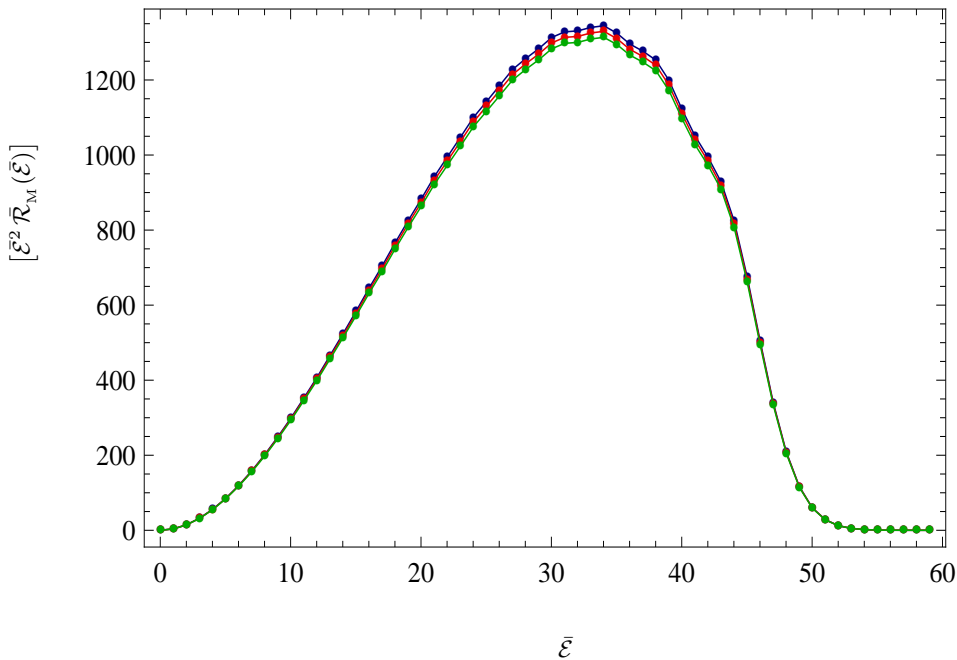}}
\vskip -170 pt \hskip -375 pt 
\rotatebox{90}{$\l[{\bar {\cal E}}^2\, 
{\bar {\cal R}}_{_{\rm M}}({\bar {\cal E}})\r]$}
\vskip 138 pt
\hskip 35 pt ${\bar {\cal E}}$ 
\vskip 5 pt
\caption{The transition probability rate of the rotating Unruh-DeWitt 
detector that is coupled to a massless scalar field governed by the 
modified dispersion relation~(\ref{eq:mdr}), with $\alpha=-1$ and
$\beta=0$.
The blue, the red and the green dots denote the numerical results 
corresponding to the following three values of the quantity $(\sigma\, 
\Omega)=0.325$, $0.350$ and $0.375$, that we had worked with in the 
previous figure.
As in the last figure, the curves simply link the dots. 
We have set $\bkp$ to be $50$, and it should be stressed that this is 
an extremely small value for $\bkp$.
For such a value, as is evident, the modifications to the standard result 
(cf. Fig.~\ref{fig:rd-sc}) due to the sub-luminal dispersion relation 
prove to be substantial.
Actually, one has to work with reasonably large and more realistic 
values of $\bkp$ such that, say, $\l(\bkp/{\bar {\cal E}}\r)> 10^{10}$.
However, numerically, it proves to be difficult to sum the contributions
in the expression~(\ref{eq:tpr-rd-psm-sbldr}) up to such large values of 
$\bkp$.
We believe that it would be reasonable to conclude that the modifications 
to standard result due to the sub-luminal dispersion relation can be expected 
to be much smaller if we assume $\bkp$ to be sufficiently large.
Nevertheless, our analysis unambiguously points to the fact that, as is 
known to occur in other contexts, a sub-luminal dispersion relation  
modifies the standard result considerably more than a similar 
super-luminal dispersion relation.}\label{fig:rd-psm}
\end{center}
\end{figure}

When we plot the result~(\ref{eq:tpr-rd-psm-spldr}) for the response
of the rotating detector when it is coupled to a field that is governed 
by a super-luminal dispersion relation, we find that it does not differ 
from the standard result (as plotted in Fig.~\ref{fig:rd-sc}) even for 
an unnaturally small value of $\bkp$ such that $\l(\bkp/{\bar {\cal E}}\r)
\simeq 10$.
In other words, super-luminal dispersion relations do not alter 
the conventional result to any extent.
It is worthwhile pointing out that similar conclusions have been arrived 
at earlier in the context of black holes as well as inflationary cosmology.
In these contexts, it has been shown that Hawking radiation and 
the inflationary perturbation spectra remain unaffected due to 
super-luminal modifications to the conventional, linear, dispersion 
relation~\cite{tpp-bhe,tpp-ic}.
In Fig.~\ref{fig:rd-psm}, we have plotted the transition probability 
rate~(\ref{eq:tpr-rd-psm-sbldr}) of the rotating Unruh-DeWitt detector
corresponding to the sub-luminal dispersion relation that we have
considered.
Again, we have plotted the result for a rather small value of $\bkp=50$.
It is clear from the figure that the sub-luminal dispersion relation can 
lead to substantial modifications to the standard result.
We believe that the modifications from the standard result will be 
considerably smaller (than exhibited in the figure) for much larger 
and more realistic values of $\bkp$ such that, say, $\l(\bkp/{\bar 
{\cal E}}\r) > 10^{10}$.


\subsection{The case of the super-luminal, quartic dispersion relation}

In the previous sub-section, we had considered the response of the
rotating detector when the scalar field is governed by a quadratic
dispersion relation (i.e. as described by Eq.~(\ref{eq:mdr}), 
{\it but with $\beta$ set to zero}).\/ 
As we have emphasized earlier, the result~(\ref{eq:tpr-sc-rt-msev}) 
holds for {\it any}\/ dispersion relation.
In this sub-section, we shall briefly discuss the situation involving
a higher order dispersion relation, viz. the case when $\beta\ne0$.
We have already seen that, even in the quadratic case, a sub-luminal 
dispersion relation leads to substantial deviations from the standard 
result.
Therefore, we shall restrict ourselves to the super-luminal, quartic, 
dispersion relation, and examine if it leads to any significant 
modifications, in contrast to the quadratic case.

When $\alpha=\beta=1$, in the cylindrical coordinates, the quartic 
dispersion relation~(\ref{eq:mdr}) is given by  
\beq
\omega
=\l(q^{2}+k_{z}^{2}\r)^{1/2}\,
\l[1+ \l(\f{1}{\kp^2}\r)\, \l(q^{2}+k_{z}^{2}\r)
+\l(\frac{1}{\kp^4}\r)\, \l(q^{2}+k_{z}^{2}\r)^{2}\r] ^{1/2}.
\eeq
In this case, Eq.~(\ref{eq:r-eq}) proves to be cubic in $k_{z}^2$.
We find that it admits one positive, and two imaginary roots 
for~$k_{z}^{2}$.
The positive root leads to
\beq
\kappa_{z} = \pm \l(\delta^{2}_{\rm c} - q^2\r)^{1/2},
\eeq
where $\delta_{\rm c}=({\bar \delta}_{\rm c}/\sigma)$, with 
${\bar \delta}_{\rm c}^{2}$ given by
\bea
\!\!\!\!\!\!\!
{\bar \delta}_{\rm c}^{2}
&=& -\, \l(\frac{\bkp^2}{3}\r)
+\l(\f{(2^{4/3}\,\bkp^2/3)}{\l(-7 -27\, (\sigma\, \Omega\,
{\bar \delta}/\bkp)^2 
+ 3^{3/2}\, \l[3+14\;(\sigma\, \Omega\, {\bar \delta}/\bkp)^2
+27\; (\sigma\, \Omega\, {\bar \delta}/\bkp)^4\r]^{1/2}\r)^{1/3}}\r)\nn\\
& &\qquad\qquad
-\l(\f{\bkp^2}{3\times 2^{1/3}}\r)\l(-7 -27\; (\sigma\, \Omega\,
{\bar \delta}/\bkp)^2 
+ 3^{3/2}\, \l[3+14\;(\sigma\, \Omega\, {\bar \delta}/\bkp)^2
+27\; (\sigma\, \Omega\, {\bar \delta}/\bkp)^4\r]^{1/2}\r)^{1/3}.
\eea
Also, for $\kappa_{z}$ to be real, we require that $q\le \delta_{\rm c}$. 
Therefore, the dimensionless transition probability rate of the detector 
is given by
\beq
{\bar {\cal R}}_{_{\rm M}}({\bar {\cal E}})
=\l(\f{\sigma}{2\,\pi\,\gamma}\r)\
\sum^{\infty}_{m\ge{\bar {\cal E}}}\, 
\l[1+\l(\f{2\,\delta_{\rm c}^2}{\kp^2}\r)
+\l(\frac{3\, \delta_{\rm c}^4}{\kp^4}\r)\r]^{-1}\;
\int\limits_{0}^{\delta_{\rm c}}\; dq\; q\;
\l(\f{J_{m}^{2}(q\,\sigma)}{\l(\delta_{\rm c}^{2}-q^{2}\r)^{1/2}}\r)
\eeq
and, upon performing the integral over $q$, we arrive at the result
\bea
{\bar {\cal R}}_{_{\rm M}}({\bar {\cal E}})
&=&\l(\f{1}{2\,\pi\,\gamma}\r)\
\sum^{\infty}_{m\ge {\bar {\cal E}}}\,
\l(\f{\l({\bar \delta}_{\rm c}\r)^{(2\,m+1)}}{\Gamma\l(2\,m+2\r)}\r)\; 
\l[1+\l(\f{2\, {\bar \delta}_{\rm c}^2}{\bkp^2}\r)
+\l(\f{3\, {\bar \delta}_{\rm c}^4}{\bkp^4}\r)\r]^{-1}\nn\\
& &\qquad\qquad\qquad\qquad\qquad\qquad\qquad\quad\times\;
_1F_2\l[[m+(1/2)];[m+(3/2)],(2\,m+1);
-\l({\bar \delta}_{\rm c}\r)^2\r].
\eea  
Note that, as $\bkp\to\infty$, ${\bar \delta}_{\rm c}\to (\sigma\,
\Omega\, {\bar \delta})$, and the above expression reduces to the 
standard result~(\ref{eq:tpr-sc-rt-smfv}), as expected.
Further, as in the case of the quadratic, super-luminal dispersion 
relation that we had considered in the last sub-section, we find 
that the above result hardly differs from the standard result even
for a rather small value of $\bkp$.


\section{The case of the rotating detector in the presence 
of a boundary}\label{sec:rd-wb}

We shall now consider the response of the rotating detector in the 
presence of an additional boundary condition that is imposed on the 
scalar field on a cylindrical surface in flat spacetime.
Due to symmetry of the problem, in this case too, it proves to be 
more convenient to work in the cylindrical coordinates, as we did 
in the last section.

Consider the time-like Killing vector associated with an observer who 
is rotating with an angular velocity $\Omega$ in flat spacetime.
Notice that the Killing vector becomes space-like for radii greater
than $\rho_{_{\rm SL}}=(1/\Omega)$.
As a result, it was argued that one has to impose a boundary condition
on the quantum field at a radius $\rho<\rho_{_{\rm SL}}$ when evaluating 
the response of the rotating detector~\cite{davies-1996}. 
Curiously, in the presence of such a boundary, it was found that the 
rotating detector coupled to the standard scalar field ceases to respond.
It is then interesting to examine whether this result holds true even
when we assume that the scalar field is governed by a modified dispersion 
relation. 

In the cylindrical coordinates, along the rotating
trajectory~(\ref{eq:rt-cyc}), the Wightman function corresponding to a 
scalar field that is assumed to vanish at, say, 
$\rho=a\;(<\rho_{_{\rm SL}})$, can be expressed as a sum over the normal
modes of the field as follows~\cite{davies-1996}:
\beq
G^{+}_{_{\rm M}}[\tx(\tau),\tx(\tau')]
=G^{+}_{_{\rm M}}(u)
=\sum\limits_{m=-\infty}^{\infty}\; \sum\limits_{n=1}^{\infty}\; 
\int\limits_{-\infty}^{\infty}\; \f{dk_{z}}{\l(2\,\pi\r)^{2}\,
\l(2\,\omega\r)}\;
\l[{\cal N}\, J_{m}(\xi_{mn}\,\sigma/a)\r]^{2}\; 
e^{-i\,\gamma\,\l(\omega-m\,\Omega\r)\, u},
\label{eq:cgfn-rt}
\eeq
where $\xi_{mn}$ denotes the $n$th zero of the Bessel function 
$J_{m}(\xi_{mn}\,\sigma/a)$, while ${\cal N}$ is a normalization 
constant that is given by
\beq
{\cal N}=\l(\frac{\sqrt{2}}{a\,\vert J_{m+1}(\xi_{mn})\vert}\r).
\eeq
Note that, as in the case without a boundary, $m$ is a real integer, 
whereas $k_{z}$ is a continuous real number. 
But, due to the imposition of the boundary condition at $\rho=a$, the
spectrum of the radial modes is now discrete, and is described by the
positive integer~$n$.
It should be pointed out that the expression~(\ref{eq:cgfn-rt}) is 
in fact valid for any dispersion relation, with $\omega$ suitably 
related to the quantities $\xi_{mn}$ and $k_{z}$.
For instance, in the case of the modified dispersion 
relation~(\ref{eq:mdr}) with $\beta$ set to zero, the quantity $\omega$ 
is given by
\beq
\omega
=\l[\l(\xi_{mn}/a\r)^2 + k_{z}^2\r]^{1/2}\;
\l(1+\l(\frac{\alpha}{\kp^{2}}\r)\;
\l[\l(\xi_{mn}/a\r)^2+k_{z}^{2}\r]\r)^{1/2},\label{eq:wrc}
\eeq
where, it is evident that, while the overall factor corresponds to the 
standard, linear, dispersion relation, the term involving $\alpha$
within the brackets arises due to the modifications to it.
Since the Wightman function depends only $u$, the transition probability 
rate of the detector simplifies to
\beq
{\cal R}_{_{\rm M}}({\cal E})
=\sum\limits_{m=-\infty}^{\infty}\; \sum\limits_{n=1}^{\infty}\;
\int\limits_{-\infty}^{\infty}\; \f{dk_{z}}{\l(2\,\pi\r)
\l(2\,\omega\r)}\; 
\l[{\cal N}\, J_{m}(\xi_{mn}\,\sigma/a)\r]^{2}\; 
\delta^{(1)}\l[{\cal E}+\gamma\, \l(\omega-m\,\Omega\r)\r].
\eeq
For exactly the same reasons that we had presented in the last 
section, the delta function in the above expression can be 
non-zero only when $m>0$.
In fact, the detector will respond only under the condition
\beq
\l(m\,\Omega\r) 
> \l(\xi_{m1}/a\r)\; \l[1 
+ \alpha\, \l(\xi_{m1}/\kp\, a\r)^2\r]^{1/2},
\eeq
where the right hand side is the lowest possible value of $\omega$ 
corresponding to $n=1$ and $k_{z}=0$.
However, from the properties of the Bessel function, it is known
that $\xi_{mn}>m$, for all $m$ and $n$ (see, for instance, 
Ref.~\cite{abramowitz-1965}).
Therefore, {\it when $\alpha$ is positive},\/ $(\Omega\, a)$ has to 
be greater than unity, if the rotating detector has to respond. 
But, this is not possible since we have assumed that the boundary at 
$a$ is located {\it inside}\/ the static limit $\rho_{_{\rm SL}}
=(1/\Omega)$.
This is exactly the same conclusion that one arrives at in the standard 
case~\cite{crispino-2008,davies-1996}.

In fact, it is straightforward to see that the above conclusion would 
apply for all super-luminal dispersion relations.
However, it seems that, under the same conditions, the rotating detector 
would be excited by a certain range of modes if we consider the scalar 
field to be described by a sub-luminal (such as, when $\alpha <0$) 
dispersion relation!
Actually, this feature is rather easy to understand.
Consider a frequency, say, $\omega$, associated with a mode through the
linear dispersion relation. 
Evidently, a super-luminal dispersion relation raises the energy of all
the modes, while the sub-luminal dispersion relation lowers it.
Therefore, if the interaction of the detector with a standard field does 
not excite a particular mode of the quantum field, clearly, the mode 
is unlikely to be excited if its energy has been raised further, as in a 
super-luminal dispersion relation.
However, the motion of the detector mode may be able to excite a mode 
of the field, if the energy of certain modes are lowered when compared 
to the standard case, as the sub-luminal dispersion relation does.


\section{Discussion}\label{sec:discussion}

In this work, we have studied the response of a rotating Unruh-DeWitt
detector that is coupled to a massless scalar field which is described 
by a non-linear dispersion relation in flat spacetime. 
Unlike, say, the case of the uniformly accelerating 
detector~\cite{rinaldi-2009,campo-2010}, defining the transition 
probability rate of the rotating detector does not lead to any 
difficulties and, we find that, it can be defined in exactly the 
manner as in the standard case.
Since it seems to be impossible to evaluate the modified Wightman function 
in a closed form, we had adopted a new method to evaluate the response of 
the rotating detector. 
However, as the transition probability rate for the rotating detector 
proves to be difficult to evaluate analytically, we had to calculate the
response of the detector numerically.
We have shown that the response of the rotating detector can be computed 
{\it exactly}\/ (albeit, numerically) even when it is coupled to a field 
that is governed by a non-linear dispersion relation. 
We have illustrated that the Planck scale modifications due to the 
non-linear dispersion relation turn out to be extremely negligible 
when the dispersion relation is super-luminal.
However, we find that there can be a reasonable extent of changes to
the standard results when one considers a sub-luminal dispersion
relation.
In addition, we have also considered the response of the rotating detector 
when the field is subjected to a boundary condition on a cylindrical surface 
located inside the static limit in the rotating frame. 
It is known that, in the standard case, the rotating detector fails to 
respond in such a situation~\cite{davies-1996}. 
We have shown that the null result remains true even for the case with 
a modified dispersion relation, provided the dispersion relation is a 
super-luminal one.
 
As we have discussed earlier, scalar fields that are governed by non-linear
dispersion relations in flat spacetime are described by actions that break 
Lorentz invariance.
The lack of Lorentz invariance implies that not all inertial frames 
are equivalent, and there exists a special inertial frame with respect 
to which the dispersion relation describing the field has been
specified~\cite{jacobson-2001,lvm-reviews}.
Therefore, the calculation that we have carried out is applicable in 
this special frame.  
Also, as we had pointed out, when the scalar field breaks Lorentz invariance, 
the modified Wightman function, in general, ceases to be invariant under time
translations in non-inertial frames that are integral curves of time-like 
Killing vector fields.
However, if the modified theory possess rotational invariance, then, possibly, 
the corresponding Wightman function can be expected to be time translational 
invariant along trajectories that respect this symmetry, as it occurs in 
the case of the rotating coordinates. 
It is interesting to examine whether there also exist other non-inertial
trajectories that possess such a property.
We are currently investigating such issues.


\appendix
\section*{Appendix~A:~Accuracy of the numerical 
computations}\label{app:comparison}

\begin{figure}[!htb]
\vskip 15pt
\begin{center}
\resizebox{330pt}{220pt}{\includegraphics{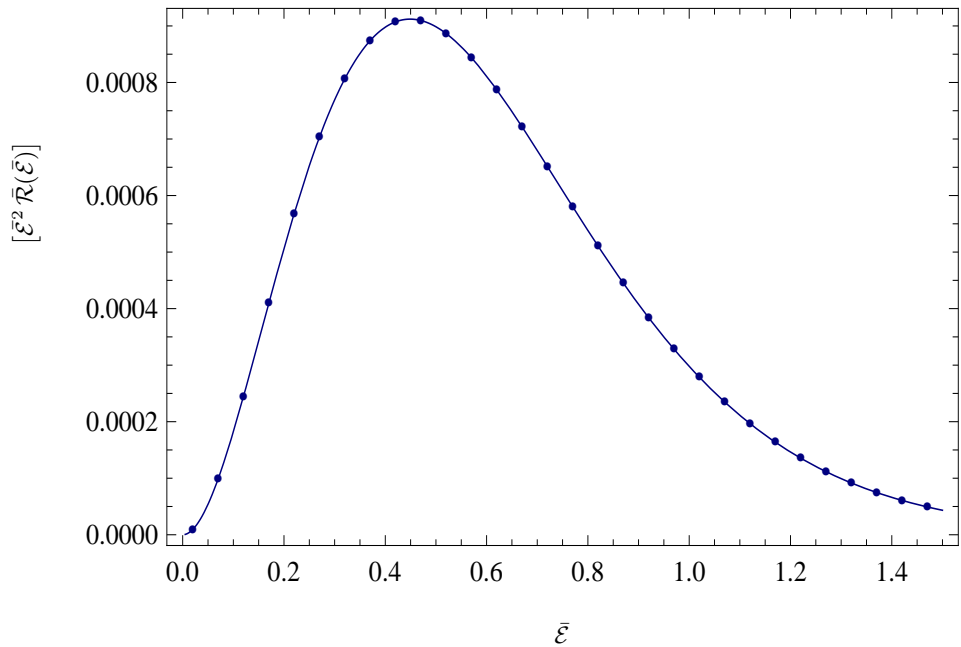}}
\vskip -170 pt \hskip -375 pt 
\rotatebox{90}{$\l[{\bar {\cal E}}^2\, 
{\bar {\cal R}}({\bar {\cal E}})\r]$}
\vskip 140 pt 
\hskip 30 pt ${\bar {\cal E}}$ 
\vskip 5 pt 
\caption{The numerical and the analytical results for the transition 
probability rate of the uniformly accelerated detector that is coupled 
to a scalar field which is described by the standard, linear dispersion 
relation have been plotted.
While the solid blue curve denotes the analytical result~(\ref{eq:dr-uam}),
the dots lying on the curves denote the corresponding results from our 
numerical computation.
Note that the plot does not explicitly depend on the acceleration
parameter $g$.
Evidently, the numerical and the analytical results are in good 
agreement.}
\label{fig:uam}
\end{center}
\end{figure}

Since the response of the rotating detector can not be evaluated 
analytically, we had initially computed the response by numerically 
evaluating the integral~(\ref{eq:tpr}) along the rotating trajectory, 
with the Wightman function being given by Eq.~(\ref{eq:mgfn-sc-rt}).
In order to illustrate the accuracy of our numerical procedure to
evaluate the integral, in this appendix, we shall compare the results 
for the transition probability rate from our numerical code with the 
analytical result that is available for the uniformly accelerated 
motion~\cite{unruh-1976,dewitt-1979}.
We shall carry out the comparison for the case wherein the scalar field 
is described by the standard, linear dispersion relation.
In such a case, the dimensionless transition probability rate of a 
detector that is moving along the trajectory
\beq
\tx(\tau)=g^{-1}\, 
\l[\sinh\,(g\,\tau),\,\cosh\,(g\,\tau),\,0,\,0\r],
\eeq
where $g$ is the proper acceleration in the comoving frame, is well 
known to be (see, for instance, Ref.~\cite{sriram-2002})
\beq
{\bar {\cal R}}({\bar {\cal E}})
=\l[{\cal R}({\bar {\cal E}})/g\r]
=\l(\f{1}{2\, \pi}\r)\, 
\l(\f{\bar {\cal E}}{e^{(2\,\pi\,{\bar {\cal E}})}-1}\r),
\label{eq:dr-uam}
\eeq
with ${\bar {\cal E}}=({\cal E}/g)$.
In Fig.~\ref{fig:uam}, we have plotted the numerical as well as the above 
analytical result.
It is obvious from the plot that the numerical result matches the 
analytical one quite well.

  
\end{document}